\documentclass[sigconf,nonacm]{acmart}

\AtBeginDocument{%
  \providecommand\BibTeX{{%
    \normalfont B\kern-0.5em{\scshape i\kern-0.25em b}\kern-0.8em\TeX}}}

\begin{document}

\title{Atomized Search Length: Beyond User Models}

\author{John Alex}
\affiliation{
  \institution{Google}
  \streetaddress{1600 Amphitheatre Parkway}
  \city{Mountain View}
  \state{California}
  \country{USA}
  \postcode{43017-6221}
}\email{jpalex@google.com}
\orcid{1234-5678-9012}
\author{Keith Hall}
\affiliation{
  \institution{Google}
  \streetaddress{1600 Amphitheatre Parkway}
  \city{Mountain View}
  \state{California}
  \country{USA}
  \postcode{43017-6221}
}\email{kbhall@google.com}
\orcid{0000-0002-0977-6572}
\author{Donald Metzler}
\affiliation{
  \institution{Google}
  \streetaddress{1600 Amphitheatre Parkway}
  \city{Mountain View}
  \state{California}
  \country{USA}
  \postcode{43017-6221}
}\email{metzler@google.com}

\renewcommand{\shortauthors}{Alex et al.}

\begin{abstract}

We argue that current IR metrics, modeled on optimizing user experience, measure too narrow a portion of the IR space.  If IR systems are weak, these metrics undersample or completely filter out the deeper documents that need improvement.  If IR systems are relatively strong, these metrics undersample deeper relevant documents that could underpin even stronger IR systems, ones that could present content from tens or hundreds of relevant documents in a user-digestible hierarchy or text summary. We reanalyze over 70 TREC tracks from the past 28 years, showing that roughly half undersample top ranked documents and nearly all undersample tail documents.  We show that in the 2020 Deep Learning tracks, neural systems were actually near-optimal at top-ranked documents, compared to only modest gains over BM25 on tail documents. Our analysis is based on a simple new systems-oriented metric, 'atomized search length', which is capable of accurately and evenly measuring all relevant documents at any depth.

\end{abstract}



\settopmatter{printfolios=true}

\maketitle

\section{Introduction}
Evaluation has played a critical role in the evolution, advancement, and wide-spread success of information retrieval systems over time. The establishment of evaluation forums like TREC~\footnote{\url{https://trec.nist.gov/}}, CLEF~\footnote{\url{https://clef-initiative.eu}}, NCTIR~\footnote{\url{http://research.nii.ac.jp/ntcir/}}, and FIRE~\footnote{\url{http://fire.irsi.res.in/}} have left an indelible mark on the community.

Benchmarks, which consist of a task, data set, and a target metric, have served as the primary force for algorithmic improvements over the years. Such benchmarks spurred the evolution from Boolean retrieval models to vector space models~\cite{saltonCACM75} to probabilistic models~\cite{Fuhr18} to language models~\cite{ponteSIGIR98} to learning to rank models~\cite{liuFNTIR09} to today's neural IR models~\cite{mitraFNTIR18}. Indeed, algorithmic improvements are tightly intertwined with evaluation metrics, as algorithmic superiority is often demonstrated via empirical evaluations. Despite this culture of algorithmic advancement via empirical evaluation, no IR technique has decisively outperformed BM25 over the past 20 years (see also~\cite{armstrongCIKM09}).  Now that neural nets (themselves developed outside IR) have finally done so, we wonder if the IR community will find itself stuck yet again, perhaps relegated to inheriting improvements from the next, ever-larger, large-LM system (such as ~\cite{raffel2020exploring}).

As algorithms have advanced, so too have evaluation metrics. While classic evaluation metrics primarily focused on system performance, there has been more interest lately in developing metrics backed by realistic user models~\cite{chapelleCIKM09,moffatTOIS08}. Metrics that correlate with user satisfaction are critically important from a product and usability standpoint, but may not be the best from the perspective of building robust and generalizable search algorithms, as it relies on an unproven assumption: that a globally optimal system (all relevant documents at the top of the list) can be achieved over time through a sequence of incrementally better systems, where each increment is required to manifest user-visible improvements. For example, a system with a relevant documents at ranks 30 and 1000 could first improve the rank 30 document to rank 2 (user-visible) while also improving the rank 1000 document to rank 500 (not user-visible), laying the groundwork for a successor system to raise the rank 500 document to a user-visible rank.  We believe this assumption does not always hold, and may have contributed to the field being stuck on BM25 for so many years.

Exclusive focus on top (user-visible) documents introduces problems for both weak and strong systems. For weak systems, user metrics are looking in the wrong place; improvements would manifest far deeper than where users would look.  We show in Section \ref{headroom} that a majority of historical TREC tracks have fallen into this category.  At the other extreme, for very strong systems, user metrics' lack of interpretability hides how saturated the metrics are. In Section \ref{neural} we show how precision-based metrics hid both how close to optimal neural systems have become (looking at the top 10 relevant documents) and how far from optimal they remain (looking at all relevant documents). The latter is important for next-generation IR systems which could present content from tens or hundred of relevant documents -- not via individual links, but perhaps via text summary generation and topic-based grouping. This would be particularly useful when we consider topics beyond simple factoid-type queries, where documents can overlap in their information, contradict each other, and describe different facets at different levels of granularity\footnote{Our metric does not try to resolve these overlaps, but does reward a retrieval system placing a large number of such relevant documents at the top, a precondition for such resolution.}. We wish to enable improvements on both types of systems by expanding our vision to improved ranking of all documents, not just user-visible ones today.

This paper specifically focuses on how to holistically and robustly measure the performance of information retrieval systems. We do this by revisiting and re-imagining the classic expected search length metric~\cite{Cooper68}. We propose a novel variant that provides more robust, intuitive, and holistic measures of a system's effectiveness. This is essentially accomplished by removing all weighting, scaling, and coupling of per-document measures.

Given the newly proposed metrics, we then revisit the past 28 years of TREC evaluations through the lens of these new metrics. We show that effectiveness, as measured with user-based metrics, is too poor to detect improved performance for approximately half of the tracks. We also demonstrate that nearly all tracks have significant headroom for retrieving additional relevant documents, to support more information-rich presentation. We compare the best neural vs. BM25 systems in the 2020 Deep Learning tracks and show that neural systems are far closer to optimal than previously realized for the top ten relevant documents, but still quite far from optimal for the rest.

To summarize, the primary contributions of this work are:
\begin{itemize}
    \item A novel retrieval metric, namely \emph{atomized search length}, that is inspired by the classic 'expected search length' metric. This is a systems-focused metric, not user-centric, and is meant to provide a robust and holistic measure of a system's performance.
    \item A detailed empirical analysis of the past 28 years of TREC evaluations through the lens of the new metric. The analysis shows that the top documents in the majority of tracks are outside the realm of user metrics (e.g. in ranks 20+) and therefore poorly measured, even when their MAP scores seem strong. In other cases, our metric shows systems to be very strong, even approaching optimality, while the corresponding user metrics (due to their uninterpretability) hide this fact. 
    \item The empirical analysis also shows that there were many forks in the road, historically, where our holistic metric would have chosen a different system ranking than user-focused metrics.  With these different choices in the past, the field might have identified a successor to BM25; and with different choices in the future, it might yet contribute to future large LM work.
\end{itemize}

\section{Related Work}
Many evaluation metrics have been proposed for IR.  Most surveys only discuss the most popular ones, but \cite{Demartini06} provides a more thorough list of 44 proposed metrics from 1960-2000.  It covers nearly all the metrics frequently used today: Precision@N, Recall, Mean Average Precision (MAP), and R-precision. The popular metrics NDCG \cite{Jarvelin02} and MRR \cite{Voorhees00} were also introduced at roughly the same time. 

Our definition of atomized search length can be thought of as a multi-document extension of 'expected search length' as defined in \cite{Cooper68}.  We resolve crucial practical details about how to use it - earlier work did not discuss which relevant documents to include nor how to include multiple relevant documents at once; nor how to handle relevant documents beyond the top-$n$.   We chose not to use the normalization step from the original paper - dividing the calculated search length by the search length of a random sampling of the corpus. This idea was perhaps based on small corpora where a random sampling could be competitive. With modern document collections numbering in the millions of documents, the original normalization is less compelling for a benchmark. 

This paper also relates to three more abstract strands of thought in IR.

First, the primacy of user experience is widely accepted in the field. The basic form of an IR metrics paper is to claim that it better models user behavior.  In a discussion of overall IR philosophy, \cite{Voorhees02} states ``since the goal is to determine how well a retrieval system meets the information needs of users, user-based evaluation would seem to be much preferable over system evaluation: it is a much more direct measure of the overall goal.'' (and the only reason for the 'seem' wording is that user-based evaluation is difficult).   In this paper we wish to add a seemingly small tweak: we also wish to consider the user experience of future systems.  That is, gains that are not user-visible now may lay the groundwork for user-visible gains in successor systems whether because deeper results finally rise high enough in future systems for users to see, or because future IR systems scale up the number of top documents they present information from, e.g. by presenting a summary of the top $n$ documents.  With that seemingly minor tweak, we effectively decouple ourselves from real-world user behavior and all the measurement challenges therein.  Considering our proposed metric as embodying a user model seems like a poor fit: our user would have infinite perseverance, would penalize each irrelevant document as equally bad, and would count that penalty multiple times - once for every relevant document eventually found beyond it. Yet we will argue that optimizing for this measure will, in the end, benefit end users in a way that optimizing for user metrics cannot.

Second, in a discussion of the overall philosophy of IR, \cite{Voorhees02} states ``the absolute score of an evaluation measure for some retrieval run is not meaningful in isolation.''  And indeed, claims of the absolute quality of IR systems are extremely scarce.   But we find this guideline’s wording overly strong.  We claim that an absolute score, in isolation, can tell us how close a system came to optimal performance on a given document collection.  In addition, while not the focus of this paper, we think that surely a score on one document collection provides at least \emph{some} signal about the system’s performance in at least some other settings (different queries and/or document collections).

Third, the minimal size of an improvement from a technique has long been judged by the 30-year-old \emph{ad hoc} parameters of \cite{SparckJones77} - 5\%-10\% improvement in system metrics, where the actual meaning of an improved metric remained opaque (e.g. what does a .1 improvement in MAP really mean?).  The interpretability of the metric we introduce in this paper, atomized search length, provides an additional way to decide if a given technique is worthwhile, beyond opaque percent-improvements or significance testing thresholds.  Lowering our metric by 5, for example, means that all relevant documents are 5 ranks higher, on average -- a more grounded and interpretable description.

\section{Methods}
This section describes the \emph{atomized search length} (ASL) metric. The name describes two key aspects of the metric: 
\begin{itemize}
    \item \textbf{Atomized:} each relevant document is measured independently of the others.
    \item \textbf{Search Length:} each document is measured (and aggregated) based on its search length instead of its precision. 
\end{itemize}

A simple interpretation is that ASL measures the average rank of all relevant documents, with the tweak that rank for each relevant document is calculated assuming all other relevant documents do not exist.

\subsection{Atomized Search Length} \label{def2}
The core of the new metric is a per-(query, relevant document) measure 
\begin{equation}
  ASL_{q,d} = irrel_{q,d} + 1
\end{equation}
Where $irrel_{q,d}$ is the number of irrelevant documents above relevant document $d$ for query $q$. We term this 'atomized' because it does not depend on the position of other relevant documents (contrast with Equation~\ref{eq:ap_perdoc} for MAP). We compute it by removing the other relevant documents from the rank:
\begin{equation}  \label{eq:asl_perdoc}
   ASL_{q,d} = rank_{q,d} - rel_{q,d}
 \end{equation}
Where $rank_{q,d}$ is the rank of the document (starting at 1) and $rel_{q,d}$ is number of relevant documents above the document. 

If a relevant document is not within the set of documents predicted by the system under test, we define its ASL as:
\begin{equation}
    ASL_(q,d) = irrel_{q}
\end{equation}
where $irrel_{q}$ is the total number of irrelevant documents among all predicted documents for $q$. 

We define ASL over a set of queries $Q$ as averaging per-document values:
\begin{equation} \label{eq:asl_average}
   ASL = \frac{1}{Q} \sum_{q=1}^{Q} \frac{1}{n_q} \sum_{d=1}^{n_q}  ASL_{q, d}
\end{equation}
where $n_q$ is the number of relevant documents for query $q$.
We propose this metric as a replacement for MAP, because both are metrics computed over all documents. In this paper we will report all ASL values (including averages) rounded to the nearest integer.  There is a simple interpretation to two systems with ASLs that differ by less than 1.0: their gold documents are, on average, within 1 rank of each other. This minor difference may be important to report in certain edge cases but we have found in practice, in the systems we evaluate below, that the differences are much larger.

Although the primary focus of this paper is on measuring all documents, we can also define the metric over subsets of relevant documents, in particular the first $n$:
  \begin{equation}
   ASL@g_{1-n} = \frac{1}{Q} \sum_{q=1}^{Q} \frac{1}{n} \sum_{d=1}^{n}  ASL_{q, d}
 \end{equation}
We propose $ASL@g_{1-1}$ as a replacement for $MRR$ and $ASL@g_{1-10}$ as a replacement for $P@20$. We note that since all ASL variants measure an explicitly defined set of relevant documents, we have no need to separately measure recall.

\subsection{Relationship to MAP}
In this section we contrast our choices above in constructing ASL with those implicit in the definition of MAP.

The per-(query, document) measure of MAP, unlike ASL, is not truly a per-document measure, because the measures of deeper documents are scaled up by the number of relevant documents above them ($rel_{q,d}$). So instead of subtracting $rel_{q,d}$ from the rank, MAP multiplies by it:
\begin{equation}\label{eq:ap_perdoc}
   P_{q,d} = (rel_{q,d} + 1) / rank_{q,d}
\end{equation}
Additionally, instead of including the rank in the numerator, MAP uses its reciprocal. This choice will turn out to be crucial below.

If a relevant document is not within the set of documents predicted by the system under test, MAP defines its P as 0. This is problematic both because 0 is not a valid rate and because it effectively estimates the rank of the missing document as infinity. One could not use an approximation like ASL does (the total number of irrelevant predicted documents), because there is no equivalent for estimating $rel_{q,d}$.

These problems come to a head, counterintuitively, at the averaging step:
\begin{align}
   AP_q &= \frac{1}{n_q} \sum_{d=1}^{n_q}  P_{q, d} \label{eq:ap_average} \\
   MAP &= \frac{1}{Q} \sum_{q=1}^{Q} AP_q \label{eq:map_average}
\end{align}

While this appears to be a simple pair of unweighted averages, we demonstrate below how they are in fact both implicitly weighted averages, because precision ($P$) is not the actual rate we want to average; $1/P$ is.  We therefore are careful to calculate ASL in $1/P$-space rather than $P$-space.  However, this necessitates all precision values being nonzero, which is not possible with the traditional formulation above. Conveniently, by decoupling our per-(query, document) metric from other relevant documents, as ASL does, we avoid this problem.

It is not immediately obvious why $1/P$ might be preferable to $P$ for averaging; we surmise that the choice, in general, of using $a/b$ vs $b/a$ has largely been governed by mathematical convenience and convention as opposed to explicit analysis.  However, it makes a profound difference when averaging rates and, in this case, we claim only one choice matches our expectations of an unweighted average.  Using the standard average of $P$, if we average the precisions of a relevant document at rank 1 and another at rank 1000, we get the precision of a document approximately at rank 2:
\begin{equation}
   (1/1 + 1/1000) / 2 \approx 1/2
\end{equation}

We have no plausible explanation for how this weighting is desirable when averaging query measures (Equation~\ref{eq:map_average}). When averaging document measures (Equation~\ref{eq:ap_average}), one might hypothesize that there is some latent quality one is truly averaging here - that in IR, perhaps the upsides of a rank 1 document far outweigh the downsides of a rank 1000 document, perhaps according to some user model; however, this is far from the simple unweighted average we aim for, and does not match this paper's goal to equally measure all documents no matter how deep they are in the ranked list. Whereas if we average the same two documents with $1/P$, the result is approximately the same as the value of a document midway between the two predictions:
\begin{equation}
   (1 + 1000) / 2 \approx 500
\end{equation}

We analyze the impact of this implicit weighting in Appendix~\ref{proof}, showing that $P$ leads to higher averages than $1/P$ because it implicitly downweights lower precisions.  One surprising implication is that MAP effectively computes the average precision of the top \textbf{57\% +-16} relevant documents, looking across all TREC tracks (see Section~\ref{data}) over time.  That is, this supposed 'holistic' measure ignores about half of the rated relevant documents.

This same argument applies to evaluating the optimality of a precision score in isolation. The top half of the precision range covers only the difference between precision of a document at rank 1 vs rank 2, while the bottom half covers ranks from 2 up through the millions (e.g., up to the size of the document collection).  The same applies to comparing per-system aggregate precisions - differences between systems with relevant documents at ranks 1 and 2 are blown up, while differences between systems with relevant documents at deep ranks are squashed. Whereas ASL provides a flat scale, as articulated in \cite{Lin21} with the perhaps theoretical caveat that unlike that work, we do not require our scale to be grounded in real-world user utility.

The derivation for ASL is thus very simple but implicitly and intentionally avoids the implicit weighting, scaling, and coupling of per-document measures noted above.

\subsection{Relationship to Other Metrics}

Atomized search length is similar to BPREF\cite{Buckley04} in that it counts the number of irrelevant documents above each  relevant document. However, it does not limit the number of measured irrelevant documents, normalize them by the total number of irrelevant documents, nor confine its calculations to judged documents only.

Precision@$n$ and recall@$n$ are much less sensitive to document rank than ASL. While all differences in a document's rank between two systems are reflected in ASL, nearly all are \emph{ignored} in precision and recall.  The only movement captured is when a document's rank crosses the $n$th-rank threshold.  To eke out information from such a narrow detector, $n$ must be chosen carefully, and we will show in Section \ref{headroom} that a reasonable value varies by at least an order of magnitude across datasets (not to mention between competing systems), whereas common practice is to set it to an arbitrary constant like 10, 20, or 100, resulting in undersampling of changes.  To characterize a system's performance using more complete information, we propose analyzing a histogram of ASLs (as in Figure~\ref{fig:one_sys}) instead of precision/recall curves.

In the next section we will analyze many TREC tracks. Some tracks have defined to task-specific metrics, but these are mostly adaptations of standard IR metrics to handle multiple types of ratings per retrieved item. For example, in the Conversational Assistance track~\cite{Dalton2020TRECC2}, NDCG is averaged over each conversation turn, while in the Decision track~\cite{Abualsaud2020OverviewOT}, NDCG is averaged across 3 types of rating (relevance, correctness and credibility).  Equivalent adaptations can be performed on ASL.

\section{Retrospective Analysis of 28 years of TREC} \label{past}
We reanalyzed 70+ TREC tracks to help demonstrate that Atomized Search Length has not only theoretical advantages over precision (including MAP) and recall but has significant practical implications on actual system evaluations.

\subsection{Data} \label{data}
In order to collect data across disparate TREC tracks, we manually corresponded per-track relevance judgments (known as 'qrels') with the per-system outputs available in the 'Past Results' section of the TREC website. We kept all tracks where: \begin{itemize}
    \item We could identify both qrels and per-system run files.
    \item All files were in expected formats.
    \item There were at least 5 run files.
\end{itemize}
We skipped the 2003 Web track because systems were run over the union of several topic sets that were intended to be evaluated separately, which would have required special-case partitioning of the data. This resulted in 76 tracks covering 4302 run identifiers and 5043 queries.

\subsection{Measuring IR Task Headroom}\label{headroom}
\begin{figure}
    \begin{center}
        \includegraphics[scale=.6]{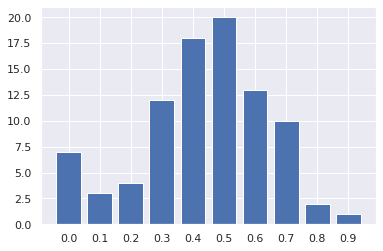}
    \end{center}
    \caption{System quality on top-ranked documents across 76 TREC tracks according to P@20.}
    \label{fig:p20}
\end{figure}

\begin{figure}
    \begin{center}
        \includegraphics[scale=.6]{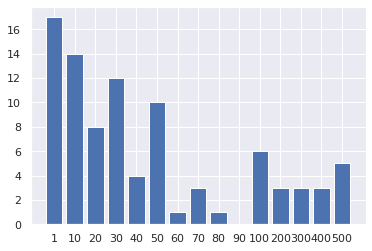}
    \end{center}
    \caption{System quality on top-ranked documents across 76 TREC tracks according to ASL@g1-10. Nearly all are beyond user-visible ranks and correspond to precisions below 0.1.}
    \label{fig:asl_g10}
\end{figure}

\begin{figure}
    \begin{center}
        \includegraphics[scale=.6]{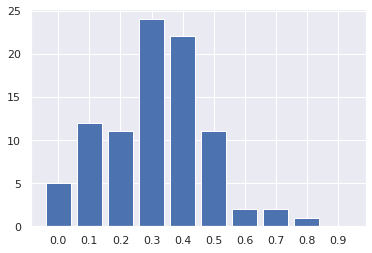}
    \end{center}
    \caption{System quality on all relevant documents across 76 TREC tracks according to MAP. }
    \label{fig:map}
\end{figure}

\begin{figure}
    \begin{center}
        \includegraphics[scale=.6]{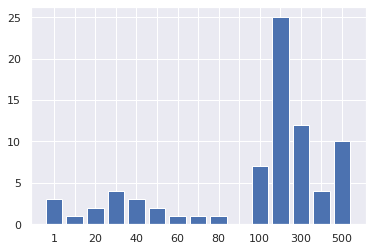}
    \end{center}
    \caption{System quality on all relevant documents across 76 TREC tracks according to ASL. The majority are beyond even rank 100, and have little correspondence with the MAP distribution in Figure \ref{fig:map}.}
    \label{fig:asl}
\end{figure}

 In this section we show that the top documents in the majority of tracks are outside the realm of user metrics and therefore poorly measured. We first consider the highest $P@20$ (averaged across all queries) for each track (Figure~\ref{fig:p20}).  From this graph we might conclude that roughly half of the track tasks are near-solved (i.e., best $P@20 \in [0.5 - 1]$)  another quarter of tasks are farther from solved (i.e., best $P@20 \in [.3, .5]$), and less than a quarter are far from solved (i.e., best $P@20 \in [0-0.3]$). However, if we look at $ASL@g1-10$ (Figure~\ref{fig:asl_g10}), again averaged across all queries, we see much worse performance. For more than half of the tracks, the average $ASL@g1-10$ is over 20, meaning there are more than 20 irrelevant documents above a given relevant document.  This means we cannot use $P@20$ (or any measure of the top 20 results) to measure improved performance from new techniques in these tracks; the measure would be undersampling. To properly sample, we would need to measure precision (or any metric) at a dynamically-choen depth comparable to what $ASL@g1-10$ indicates, which is often above 30 documents and sometimes even well into the hundreds of documents. Note these (high) ASL values are in fact lower bounds, since we use a conservative estimate of ASL for documents beyond the predicted $n$ documents per query. \footnote{TREC tracks use a varying number of results for evaluation purposes (100 and 1000 are the most common values), so for lower values (say, 50-100), we may wonder if the true value is much higher but bounded by $n$ being 100. Despite this difference, we chose to include all the tracks together in order to give our best sense of the wide range of difficulties in these tracks.} 

We also show that user-facing metrics hide headroom in well-solved tasks. In Figure~\ref{fig:asl} we examine the ASL of all relevant documents, not just the top 10.  This paints a much different picture: nearly all of the tracks are in the worst (rightmost) buckets, meaning they leave a large amount of information (relevant documents) stranded in deeper ranks.  As stated earlier, if users would not want to directly scroll through those additional documents, IR systems could perform a second level of processing on top results (e.g. grouping or summarizing) to present more information in a user-digestible way.  In contrast, the equivalent graph for MAP (Figure~\ref{fig:map}) does not show this headroom. There is a slight shift to the left compared to the P@20 graph, but it otherwise hides how deep the other documents are.  Given that nearly all the ASL values are greater than 10, we should expect MAP values to be less than .1; the fact that nearly all MAP values are so much higher is an indication of how strongly MAP down-weights deep documents.  In addition to hiding the existence of this headroom, it is unclear how one could use MAP to measure system improvements on these tasks -- while MAP does incrementally reward any documents that move up in the ranks, the magnitude of the delta is insignificant unless the document reaches a top-10 rank -- whereas ASL provides a straightforward and interpretable measure of progress.

\subsection{Reassessing Techniques: Neural vs BM25} \label{neural}
In this section we show how user metrics, being not interpretable, can occlude the fact that a task is near-solved (and perhaps worth moving on from).

\begin{table}
\caption{Technique Reanalysis in 2020 Deep Learning Tracks}
  \label{tab:dl_2020}
  \begin{tabular}{l|ccc|ccc}
    \toprule
    & \multicolumn{3}{c}{Documents} & \multicolumn{3}{c}{Passages}\\
    & BM25 & neural & RRIE & BM25 & neural & RRIE \\
    \midrule
    P@20 & 0.456 & 0.594 & 25\% & 0.517 & 0.706 & 39\% \\
    ASL@g1-10 & 14 & 5 & 69\% & 51 & 6 & 90\% \\
    \midrule
    MAP & 0.401 & 0.543 & 24\% & 0.400 & 0.572 & 29\% \\
    ASL & 39 & 27 & 32\% & 264 & 215 & 19\% \\
    \bottomrule
\end{tabular}
Top ranked and all-doc metrics for the best BM25 and neural systems, and Relative Reduction in Error (RRIE) from BM25 to neural.
\end{table}

In Table~\ref{tab:dl_2020}, we compare measurements of top ranked documents with P@20 and ASL@g1-10, and measurements of all documents with MAP and ASL. We calculate the relative reduction in error (RRIE) for P@20 and MAP as:
\begin{equation}
  \frac{(1-P_{bm25}) - (1-P_{neural})}{(1-P_{bm25})}
\end{equation}
And for ASL and ASL@g1-10:
\begin{equation}
  \frac{(ASL_{bm25} - 1) - (ASL_{neural} - 1)}{(ASL_{bm25} - 1)}
\end{equation}

ASL@g1-10 not only shows a more dramatic RRIE for neural on top ranked documents than P@20, but more importantly it is suggestively close to the optimum of 1. The improvement in Passage Retrieval was even more dramatic: the average rank went from 51 to rank 6. That is, there are only 4-5 irrelevant documents out of place (out of a corpus of millions) with respect to the first 10 relevant documents. The traditional user metrics aren't interpretable but as a general rule of thumb they do look strong: MAP in the .5's, and precision in the .6-.7 range.  What they do not suggest is that the task is approaching optimality, as ASL reveals.

Turning from top ranked documents to all documents, ASL tells a much different story than MAP, in two ways. First, ASL shows that the wins for neural systems over all documents (plain ASL) are much more modest than the wins over top documents (ASL@g1-10). This is not surprising since systems in the track were not optimized to improve deeper documents, but MAP fails to surface this information, showing a similar RRIE to the P@20 measure.  In fact, the absolute MAP scores are surprisingly close to the absolute P@20 scores, further evidence that MAP is only weakly affected by lower ranked documents, as predicted by our analysis in Appendix~\ref{proof}. Second, ASL shows that Document Retrieval is in far better shape (average rank 27) than Passage Retrieval (average rank 215), but MAP once again hides this information, giving both sides similar scores.  MAP is therefore untrustworthy both for top-ranked documents (radically understating the neural win and distance-to-optimality while ASL@g1-10 does not) and for lower-ranked documents (it ignores them, as demonstrated by its similarity to P@20 scores, whereas ASL scores tell a much different story than ASL@g1-10).

\subsection{Reordering of systems}\label{reordered}
\begin{table}
\caption{Reordering of systems across TREC tracks (all docs)}
  \label{tab:ablation_ap}
  \begin{tabular}{c|ccc}
    \toprule
    Technique & $\Delta Sort$ & Kendall & $\Delta Value$ \\
    \midrule
	Atomize P                 & 26-60\% & 0.7-0.9 & 0.3-0.6x \\
	Agg. docs within query     & 37-77\% & 0.4-0.8 & 0.03-0.4x \\
	Agg. rates across queries  & 29-78\% & 0.4-0.9 & 0.1-0.9x \\
	\midrule
	Cumulative                & 26-60\% & 0.6-0.9 & 0.004-0.06x \\
	\bottomrule
\end{tabular}

\bigskip
Measures of system reordering (and change in absolute metric values) when gradually changing our metric from Atomized Search Length (ASL) to Mean Average Precision (MAP).  Scores are presented as $(\text{mean} - \text{stddev}) - (\text{mean} + \text{stddev})$.  
\end{table}

In prior sections we argued why ASL is a better metric than existing ones from a theoretical perspective, and showed that it reveals new kinds of headroom in TREC tracks. In this section we show that ASL would have led to different technique choices, which might have allowed the field to identify a successor to BM25. Specifically, we determine whether ranking by ASL would cause any system in a given track to move significantly higher or lower versus the rest of the submitted systems in the track.

In Table~\ref{tab:ablation_ap}, the 'Cumulative' row measures the impact of ranking systems by ASL instead of MAP. This includes reporting the results in $1/P$-space instead of $P$-space (which doesn't change system ordering, but can change significance testing results).  
We report $\Delta Sort$ where $\Delta Sort_{s} = 100*\frac{|n_0-n_1|}{|S|}$ for each system $s$, where $n_0$ is number of systems better than system $s$ without the change, and $n_1$ is the number after, where 'better' means passing two significance tests: the absolute value of the metric is 10\% better, and a two-tailed Student t-test on the per-query metrics returned a probability (of the systems being identical) no greater than 5\%.  We pick the largest such change per evaluation track, then average across all tracks.  We report the Kendall rank correlation coefficient over all systems in each track, averaged over all tracks.  For more insight we also report $\Delta Value$: $\frac{new}{old}$ for the metric itself; averaged over all systems in all tracks. For this column we compare AP with $1/ASL$ to better compare their magnitudes.   We report the ranges of changes from $+/-$ one standard deviation.  

While we assumed that by changing the metrics there would be a considerable change in the ranking, we did not expect the magnitude of the effect to be quite so large - the mean delta-sort indicates that at least one system in each track jumped above or below roughly half of its competitors on average, which we argue indicates that for each of the 76 tracks, the effectiveness of at least one technique needs to be reconsidered.

We also isolate the effects of each incremental change from MAP to ASL:
\begin{itemize}
    \item \textbf{Atomize P:} Remove the cross-document coupling from per-document precision: replace Equation~\ref{eq:ap_perdoc} with the reciprocal of  Equation~\ref{eq:asl_perdoc}.
    \item \textbf{Aggregate docs within query:} Change Equation~\ref{eq:ap_average} from an arithmetic mean of $P$ to a harmonic mean.
    \item \textbf{Aggregate rates across queries:} Change Equation~\ref{eq:map_average} from an arithmetic mean of $AP$ to a harmonic mean.
\end{itemize}
The results of each ablation row are calculated relative to the intermediate metric defined in the prior row. Interestingly, each intermediate row -- representing a relatively subtle change in the math -- had a similarly large $\Delta Sort$ by itself.  These results show the importance of these seemingly minor mathematical decisions.

We also performed the same analysis with respect to top ranked documents only and found similarly large impacts to system ranking. The cumulative $\Delta Sort$ from changing P@20 to ASL@g1-10 was 40-76\% (0.4-0.8 Kendall-Tau), and changing MRR to ASL@g1 was 49-85\% (0.3-0.7 Kendall-Tau). While measuring the performance of top ranked documents is not the focus of this paper, these large $\Delta Sort$ values indicate the potential of changing even more conclusions about effective techniques from earlier tracks (especially MRR since it is the main metric of the popular MSMARCO~\footnote{\url{https://microsoft.github.io/msmarco/}} dataset) .

\subsection{Analysis Tools and Insights}
\begin{figure} 
    \begin{center}
        \includegraphics[scale=.6]{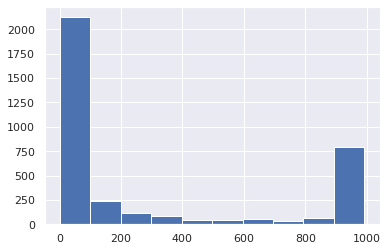}
    \end{center}
    \caption{Number of relevant documents with per-document ASL for the best 2020 Deep Learning Passage Retrieval system .}
    \label{fig:one_sys}
\end{figure}

\begin{figure} 
    \begin{center}
        \includegraphics[scale=.6]{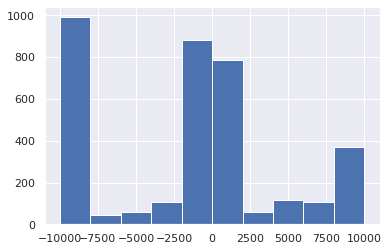}
    \end{center}
    \caption{Number of relevant documents with per-document ASL changes (best-system minus median-system) in the 2012 Web Adhoc Retrieval track}
    \label{fig:two_sys}
\end{figure}

In this section we present a more informative alternative to traditional precision/recall graphs by decomposing our ASL metric (averaged over all documents) into a distribution of per-document ASLs.  Note that per-document granularity is not possible with traditional metrics because they couple all documents for each query.  In Figure~\ref{fig:one_sys}, we show a ASL distribution across all documents for one representative system.  We generated and manually inspected the same graph for the best system in every track, and qualitatively they all had this same basic shape: a large number of documents in the leftmost bucket (which includes the optimal ASL of 1), then a falloff for larger ASL buckets, and an optional second spike at the rightmost bucket (which counts relevant documents not in the predicted set).  Depending on how well-solved the track was, the rightmost bucket ranged from being empty to being far higher than the leftmost bucket.  Note that since all measurements in TREC tracks were capped at some number of predictions (usually 100, 1000, or 10000), we cannot tell if the rightmost bucket could in fact be distributed among many deeper buckets (perhaps following the slope of the existing tail).  We find this distribution particularly surprising for the specific system in the figure (2020 Passage Retrieval) because that system is so strong: its ASL for the top 10 golds is only 5 (very close to the absolute optimum of 1).  Yet at the same time this system is apparently not capturing some key phenomenon in passage ranking: the rightmost bar indicates that there are over 750 relevant passages that \emph{each} have >900 irrelevant passages ranked above them. 

In Figure~\ref{fig:two_sys}, we extend this analysis to a pairwise comparison of systems, showing a representative distribution of per-document wins and losses between the best and median system in one track. Negative values represent a win for the best system (lower per-document ASL).  We generated and manually inspected the same graph for every track (always comparing the best to median system, as ranked by ASL) and saw the same general shape: a large number of the smallest wins and losses (the middle two buckets); a smaller and relatively balanced number of middling wins and losses; and - usually but not always - a large number of the largest wins and losses (the outer two buckets), with the number of largest wins exceeding the number of largest losses. Similar to the prior figure, we cannot tell if the outermost buckets could in fact be distributed among many deeper buckets (perhaps following the slope of the existing tails) due to the cap on the number of predictions. We found the frequent presence of these large outer buckets most surprising - we wonder if these massive shifts of 100, or 1000, or 10000 ranks might indicate the learning of larger-scale phenomena compared to large number of small deltas. We also found it surprising that the righthand half of the graph (losses of all sizes) was so substantial across all tracks, which suggests that IR improvements are an intrinsically mixed bag, with some amount of losses somehow being inevitable.  

\section{Conclusions and Future Work}
Through a novel analysis of 76 historical TREC tracks using our new metric, Atomized Search Length, we demonstrated that top relevant documents for many tasks were too far down in the result set for traditional user-centered metrics to measure accurately. On the other hand, for tracks that are well-solved under the common user scenario (i.e., scanning through ~10 documents ranked by relevance; notably the recent Deep Learning tracks), user-centered metrics hide huge headroom to surface even more relevant documents, which could support more information-rich user interfaces that organize or summarize tens or hundreds of documents (an example of work in this direction is \cite{FAIRFID21}). We also showed how Atomized Search Length removes implicit weights and couplings from traditional precision-based metrics.

Since our technique causes large changes in system sort order across TREC tracks, there are many more track-specific technique reassessments that could be done, including for the many additional TREC tracks we did not import due to special data handling needs.

We also leave the study of how other user-based metrics such as NDCG~\cite{Jarvelin02}, ERR~\cite{chapelleCIKM09}, and RBP~\cite{moffatTOIS08} might be revisited within this framework as future work. Because these metrics express more complicated functions of rank than precision, the adaptations above are not directly applicable, but the general concerns (particularly with the use of $1/P$ instead of $P$) apply.

\bibliographystyle{ACM-Reference-Format}
\bibliography{base}

\appendix
\section{Appendices}

\subsection{Severity of weighting caused by precision averaging} \label{proof}
Since any weighted average still returns a value somewhere 'in the middle' of its inputs, does it really matter that it's not perfectly unweighted? In this section we present several ways to see how much the traditional arithmetic mean of precision ($P$) is larger than the harmonic mean (the reciprocal of the average of $1/P$), which accounts for two of the three main differences between ASL and MAP (see ablation in Section~\ref{reordered}). Note the analysis in this section can be used to compare arithmetic and harmonic means in any mathematical context, not just IR.

First, we analytically show that averaging two precisions is more similar to a 'max' operation than a 'mean' - it is nearly entirely insensitive to the lower precision. We can express the average of any two precisions $P_1$ and $P_2$ as:

\begin{equation}
  avg_P = .5 (\frac{1}{SL_1} +\frac{1}{SL_2})
\end{equation}
where $SL_i$ = $1/P_i$.  If we hold $SL_1$ fixed and allow $SL_2$ to become arbitrarily large, $avg_P$ approaches $.5 * \frac{1}{SL_1}$; the corresponding output in $SL$-space is $2SL_1$. Put another way, the average of a fixed SL ($SL_1$) with an arbitrarily large one ($SL_2$), if performed in $P$-space, approaches the original value ($SL_1$) plus a small penalty (also $SL_1$). The other input, $SL_2$, controls neither of those values; it acts only as a knob to downscale the penalty even further. In contrast, a straightforward mean of $SL_1$ and $SL_2$ would grow correspondingly larger as $SL_2$ grew larger, without bound.

Second, we analytically show how an unweighted average of precision ($P_i$) values is identical to a weighted average of the equivalent $SL_i = 1/P_i$ values (which we argued is the actual thing we wish to average) with a specific choice of weights. That is, we wish to prove that the following is true for some choice of weights $w_i$:

\begin{equation}
  \frac{n}{\sum_{i=1}^{n} P_i} = \frac{\sum_{i=1}^{n} SL_i * w_i}{\sum_{i=1}^{n} w_{i}}
\end{equation}
When we specifically choose weights $w_i = 1 / SL_i$, then each right-side numerator term simplifies to 1.0, while the right-side denominator becomes the sum of all reciprocals:

\begin{equation}
  \frac{n}{\sum_{i=1}^{n} P_i} = \frac{\sum_{i=1}^{n} 1.0}{\sum_{i=1}^{n} 1/SL_i}
\end{equation}
Since by definition $P_i$ = $1 / SL_i$, the two sides are indeed identical. Thus any unweighted average of precisions is exactly the same as a weighted average of equivalent search lengths  where each search length is downweighted by its own magnitude. The input documents or queries with the highest precisions -- the ones with the lowest search lengths -- thus dominate the output, resulting in artificially higher averages. 

We can use this identity to characterize an unweighted average in $P$-space as an unweighted average in $SL$-space of a subset of the inputs - the top $n$. To calculate $n$, we sort all documents by decreasing weight (that is, by decreasing precision) and then binarize their weights - the highest-weight $n$ documents will get a binarized weight of 1, and the rest will get 0. We find a partition point $n$ such that the resulting average in $SL$-space is as close as possible to the original average in $P$-space.  When calculated over every runfile in each TREC track, and then averaged across tracks, we find that traditional MAP effectively returned the average precision of the top \textbf{57\%} +- 16 of docs.  That is, this supposed 'holistic' measure of all documents actually includes only half.

Finally, to gain more insight into the severity of the downweighting, we can swap out each weighted value for a value that results in the same average if unweighted ($w_i = 1.0$). Given input search lengths of:
   [1, 1, 1, 2,   4,   5,   10, 10000, 10000], 
an unweighted average by $P$ is equivalent to an unweighted average of these $SL$s instead:
   [1, 1, 1, 2.1, 2.7, 2.9, 3,    3.2,   3.2]. Small $SL$s move a little, while large $SL$s shrink a lot - to approximately the same destination value, nearly independent of their original magnitude. This is not surprising since most of the $SL$ range ($SL$ from 10 up to the corpus size, O(millions)) is squashed into the 0-.1 range in $P$-space.

\end{document}